\documentclass[12pt,fleqn]{article}

\usepackage{epsfig}
\usepackage{amsmath}
\usepackage{times}
\usepackage[sort&compress,comma]{natbib}
\usepackage[hang,small]{caption}

\newcommand{\Eq}[1]{Equation~(\ref{-1})}
\newcommand{\eq}[1]{Eq.~(\ref{-1})}
\newcommand{\eqs}[1]{Eqs.~(\ref{-1})}
\newcommand{\fig}[1]{Fig.~\ref{-1}}
\newcommand{\Fig}[1]{Figure~\ref{-1}}
\newcommand{\note}[1]{\marginpar{\bf -1}}

\makeatletter
\renewcommand\@biblabel[1]{#1.}
\makeatother

\title{\sffamily\textbf{A model for the evolutionary diversification of religions}}
\vspace{0.5cm}
\author{Michael Doebeli \& Iaroslav Ispolatov\\
\vspace{-2mm}\normalsize Department of Zoology and Department of Mathematics, \\
\vspace{7mm}\normalsize University of British Columbia, Vancouver B.C., Canada
V6T 1Z4} 
\vspace{10mm}
\date{\vspace{10mm}\normalsize\today}

\begin{document}

\maketitle


\newpage

\noindent {\bf\large Abstract}
\vskip 0.5 cm
We address the problem of diversification in religions by studying selection on cultural memes that colonize humans hosts. In analogy to studying the evolution of pathogens or symbionts colonizing animal hosts, we use models for host-pathogen dynamics known from theoretical epidemiology. In these models, religious memes colonize individual humans. Rates of transmission of memes between humans, i.e., transmission of cultural content, and rates of loss of memes (loss of faith) are determined by the phenotype of the cultural memes, and by interactions between hosts carrying different memes. In particular, based on the notion that religion can lead to oppression of lower classes once a religious society has reached a certain size, we assume that the rate of loss increases as the number of humans colonized by a particular meme phenotype increases. This generates frequency-dependent selection on cultural memes, and we use evolutionary theory to show that this frequency dependence can generate the emergence of coexisting clusters of different meme types. The different clusters correspond to different religions, and hence our model describes the emergence of distinct descendent religions from single ancestral religions.

\newpage

\noindent {\bf\large Introduction}
\vskip 0.5 cm

Evolution can occur whenever there are units of reproduction that produce
other such units
which inherit some characteristics of the parent units. If
the units of reproduction vary in their reproductive output, there will be
evolutionary change. ``Intellectual content'' can satisfy these simple
requirements. An idea or a theory can be viewed as a unit living in the brain
of an individual human (or animal). It can mutate within that brain, and it
can be passed on to the brains of other individuals, thereby reproducing
itself (typically with modification). For a multitude of potential reasons,
some ideas and theories are more successful at such reproduction through
transmission than others, hence there is typically differential reproductive
success. As a consequence, there is cultural evolution of intellectual content
such as ideas and theories. 

Based on the notion of ``meme'', this perspective has been very  lucidly
advocated by Richard Dawkins (\citet{dawkins76}). There is a large body of literature
on cultural evolution, but when cultural evolution is conceptualized, the
reproducing units are most often not the units of cultural content themselves,
but instead the (typically human) units of 
physical 
reproduction that carry the
cultural content. For example, 
such an 
approach has been used for models of the
evolution of language, in which the evolutionary dynamics of language is
determined by the reproductive success of individuals speaking the language
(e.g. \cite{nowak_krakauer1999}, \cite{mitch_nowak2003}). This is a very
interesting and valid approach that nevertheless does not treat the language
itself as the reproducing unit that is transmitted among suitable ``host''
individuals. In addition, the evolution of cultural diversity is often studied
by determining the ``winners'' among a preexisting set of different cultures
(e.g. \cite{diamond2005}, \cite{lim_etal2007}). This approach is roughly
equivalent to studying ``species selection'' between already established
species and foregoes the question of how diversity arose in the first place
within a single culture.  

It seems perhaps relatively easy to understand how cultural differentiation
can develop between human populations that live in isolation from each other
(e.g. on different continents). However, cultural differentiation also seems
to occur when people adopting diverging cultures live together. Some instances
of diversification in religion may serve as paradigms for such processes. For
example, the split of the protestant from the catholic church in the 16th
century occurred from within an essentially entirely catholic culture, and
despite some subsequent spatial segregation of the diverging religions (due,
among other things, to violent conflicts), the two religions essentially
coexisted since the split. It has been argued that this split was caused by a
decline in the moral authority of the catholic leadership
(\cite{tuchman1985}), i.e., by processes occurring within the catholic church
that led certain people to be susceptible to new religious ideas. Thus,
cultural evolution within the catholic church may have generated conditions
that favoured the emergence of a dissident religious strain. In a more recent example, \cite{whitehouse95} has observed an ongoing splitting off of minor sectarian movements from a mainstream religious organization in Papua New Guinea.

In this paper, we propose to model cultural diversification in religion using
techniques from evolutionary theory to describe scenarios in which the
reproducing units are religious memes, and the traits whose evolutionary
dynamics is investigated reflect the memes' religious content. Borrowing ideas
from epidemiology, our models incorporate human individuals as hosts for
religious memes. The trait values of these memes
determine their propensity of
being lost by their human hosts, as well as their success in colonizing
susceptible hosts. Our models are very simplistic, but they serve the purpose
of illustrating how cultural interactions can give rise to selection pressures
that act on cultural memes and generate religious diversity. In particular, we
believe that considering cultural content, and in particular religious
content, as the unit of reproduction it is a very useful perspective that
allows us to objectify the often acrimonious discussions between different
cultures and religions.  

\vskip 1cm
\noindent
{\bf\large Model setup}
\vskip 0.5cm

Religions are sets of ideas, statements and prescriptions of whose validity
and applicability individual humans can become convinced. Thus, individual
minds are the hosts of religious memes, which can exert considerable influence
on the behaviour of their hosts. In principle, understanding the dynamics of
religion can be achieved by understanding the interaction between religious
memes and their hosts, i.e., by understanding how religion affects not only
the behaviour of individual hosts, but also the social structure of host
populations, and how behaviour and social structure in turn affect the
transmission of religious memes among host individuals. Host populations of a
given religion are often hierarchically structured, with relatively few hosts
enjoying high social status, and many hosts enjoying fewer benefits from
adopting the given religion. As the number of host individuals adopting a
given religion grows, this social structure may give rise to unrest,
particularly in the lower social ranks. As a consequence, individuals may be
enticed to adopt alternative, ``unspoiled'' religions, which offer less
repression, and in which they can attain improved social status. For example,
it has been suggested that social unrest led to the split of the protestant
church from the catholic church in the 16th century (\cite{tuchman1985}). In
that time, political developments led to ever increasing financial needs of
the catholic church, which burdened its followers through taxation and other
means, e.g. the sale of indulgences. This in turn led to unrest and spiritual
decay and contributed to the secession of a more democratic and less
repressive religion. In other words, hosts of the  catholic meme tended to
lose that meme due to effects that the catholic memes themselves generated in
their host society. Moreover, faith-losing hosts became susceptible to a
similar, but distinct type of religious meme that promised to improve the
conditions of these hosts, probably at least in part because the new religious
meme was not very common, and hence did not have the same detrimental effects
on its hosts as the catholic meme. Of course, many other forces impinge on the
well-being of hosts of a particular religious memes. For example, common memes
may offer protection, and rare memes may suffer persecution. Nevertheless,
here we propose that mechanisms such as the ones alluded to above could cause
negative frequency-dependent selection on religious memes, and, as a
consequence, adaptive cultural diversification. 

Under the perspective of host individuals being colonized by religious memes,
it is natural to attempt a formal analysis of the evolution of religion using
epidemiological models. Such models are very well studied in the context of
disease dynamics (\cite{otto_day2007}). In the simplest case, there is only
one type of religion present, and the corresponding model describes the
dynamics of two variables, each describing a subpopulation of the total host
population: $S$ denotes the 
density 
of susceptible individuals in the host
population, i.e., individuals that are not yet hosts to the given religious
meme, and $C$ denotes the 
density 
of colonized hosts, i.e., host individuals
whose minds have adopted the given religion.  

Our analysis is based on the following ``ecological'' model for the dynamics
of susceptible and colonized hosts: 
\begin{align}
\frac{dS}{dt}&=r_S S\left(1-\frac{S+C}{K_S}\right)-\tau SC+lC\\
\frac{dC}{dt}&=r_C C\left(1-\frac{S+C}{K_C}\right) + \tau SC -lC
\end{align}
Here we have assumed that both susceptible and infected hosts grow
logistically. Thus, in the absence of religious memes, susceptible hosts have
an  intrinsic growth rate $r_S$ and grow logistically to carrying capacity
$K_S$, and in the absence of susceptibles, hosts colonized by the religious
meme have an intrinsic growth rate $r_C$ and grow logistically to carrying
capacity $K_C$. For simplicity, we assume that offspring of religious hosts
are also religious (in principle, part or all of these offspring could first
join the susceptible class), and that offspring of susceptible hosts also
belong to the susceptible class. The two 
death 
terms for susceptible and
colonized hosts are coupled by assuming that growth depends on the sum of the
two 
populations 
$S$ and $C$. In addition, susceptible hosts adopt religion,
i.e., become colonized, at a per capita rate $\tau C$ that is proportional to
the number of religious hosts. However, religious people also lose their
faith
at a per capita rate $l$ and become susceptible once again, leading to
a decrease in $C$ at a rate $lC$ and a corresponding increase in $S$. 

To introduce variability in religious memes and 
thus to allow for religious diversification, we expand this model by
making the very simplistic
assumption that memes are characterized by a 1-dimensional trait $x$, and that
$C(x)$ describes the distribution of the various religious
types. Mathematically, $C(x)dx$ is the population density of hosts colonized
by memes with values in the interval $(x,x+dx)$. To 
introduce
frequency-dependent selection on memes, we first introduce a measure of 
``overcrowding'' 
by defining, for any given meme type $x$, the function 
\begin{align}
A(x)=\int_y\alpha(x-y)C(y)dy,
\end{align}
where $\alpha(x-y)$ is a unimodal function of the form
\begin{align}
\alpha(x-y)= \alpha_0\exp\left(-\frac{1}{2}\left[\frac{\vert
    x-y\vert}{\sigma_\alpha}\right]^{b_\alpha}\right). 
\end{align}
The exponent $b_\alpha$ in $\alpha(x-y)$ is a positive real number. For example, if
$b_\alpha=2$, $\alpha(x-y)$ is a Gaussian function. Technically speaking, $A(x)$ is a convolution of the density distribution $C(y)$ with the ``kernel'' $\alpha(x-y)$. Such a convolution corresponds to a weighted sum over all densities $C(y)$, with the weights $\alpha(x-y)$. Since $\alpha(x-y)$ has a maximum at $x=y$ and decreases to 0 as the distance $\vert x-y\vert$ increases, the densities $C(y)$ of hosts colonized with meme types $y$ that are very different from the focal type $x$ have little weight, and hence matter little for calculating the quantity $A(x)$, where as the density of hosts colonized by meme types that are more similar to the focal $x$ have more weight in calculating the overcrowding $A(x)$ at $x$. In general, if the distribution
$C(x)$ is unimodal with a single maximum at $x=x_0$, then overcrowding 
$A(x)$, Eq. (3),
tends to be large for $x$ close to $x_0$, i.e., for common $x$, and,
conversely,  $A(x)$
tends  to be small for $x$ very different from $x_0$, i.e., for rare $x$.

We then assume that the per capita rate of loss of 
the religious meme, $l$, is
a function of 
overcrowding, 
$l(A(x))$, 
where $l(z)$ increases monotonically with increasing
$z$. For simplicity, we assume $l(z)=z$. This implies that the rate of loss is
high for hosts carrying religious memes $x$ for which $A(x)$ is large, whereas
the rate of loss is small for hosts carrying religious memes $x$ for which
$A(x)$ is small. Thus, hosts are more likely to lose common religious memes
than rare religious memes. As mentioned above, one rationale for this
assumption is that once a religion becomes common, the social structure may
change such that the benefits gained from adopting the religion decrease for
the majority of hosts, so that, on average, hosts of such memes become more
likely to lose faith. 


With religious variability, the differential equation for $C$ must be replaced
by a partial differential equation describing the dynamics of the distribution
$C(x)$. To model this, we assume that offspring of hosts colonized by meme $x$ on average also carry meme $x$, but with a certain probability the meme carried by the offspring undergoes a small mutation (as e.g. when children adopt religious notions that are slightly different from those of their parents). Thus, the offspring of a parent with religious meme $y$ has a probability $N_{y,\sigma_m}$ to lie in the interval $(x,x+dx)$, where $N_{y,\sigma_m}$ is a normal distribution with mean the parental type $y$ and mutational variance $\sigma_m$. With this in mind, the epidemiological dynamics for religiously variable host populations becomes
\begin{align}
\frac{dS}{dt}&=r_S
S\left(1-\frac{S+\int_xC(x)dx}{K_S}\right)-S\int_x\tau(x)C(x)dx+\int_xA(x)C(x)dx\\ 
\frac{\partial C}{\partial t}&=r_C \int_y N_{y,\sigma_m}C(y)dy-\frac{r_CC(x)\left(S+\int_xC(x)dx\right)}{K_C} +
\tau(x) SC(x) -A(x)C(x) 
\label{dynamics}
\end{align}
To include mutation at birth, we have separated the birth and death term of the logistic equation, with $r_C \int_y N_{y,\sigma_m}C(y)dy$ describing all offspring that are born to parents with all possible memes $y$ and whose meme mutated to $x$. In the death term, $\int_xC(x)dx$ is the total population size of religious hosts. Also, to 
describe transmission, we have made the additional assumption that the
transmission constant $\tau(x)$ is a function of the religious trait $x$. This function is assumed to reflect some intrinsic properties of religious memes that determine their likelihood of transmission to susceptible hosts. For example, some memes might entice their carriers to proselytize more than other memes, but such activities might come at certain costs. The function $\tau(x)$ is assumed to reflect the balance of such costs and benefits. Specifically, we assume that this function is unimodal, so that there is a unique ``optimal'' religious type  in terms of transmissibility. This introduces a stabilizing component of selection on the meme trait $x$. Specifically, we will use
the form 
\begin{align}
\tau(x)=\tau_0\exp\left(-\frac{1}{2}\left[\frac{\vert x-x_0\vert}{\sigma_\tau}\right]^{b_\tau}\right),
\end{align}
where 
the exponent $b_\tau$ is a positive real number. Note that for $b_\tau=2$,
$\tau(x)$ is a Gaussian function. The rate at which colonized hosts with meme
$x$ convince susceptible individuals of their religion is $ \tau(x) C(x)$, so
that the total per capita rate of transmission for susceptible hosts is
$\int_x\tau(x)C(x)dx$. 
Also, for colonized hosts with meme $x$ the per capita
rate of loss is $A(x)$ as described above, so that the total rate of loss is
$\int_xA(x)C(x)dx$. For simplicity, we assume $r_S=r_C=r$ and $K_S=K_C=K$ in
the sequel. 
We note that with the above assumptions, the religious trait $x$ only affects rates of loss and transmission, but it does not affect the birth and death rates of colonized hosts. Thus, selection on memes is not mediated by differential viability and/or reproductive success in the host population, but instead by differential loss and gain of memes by colonized and susceptible hosts.

\vskip 1cm
\noindent
{\bf\large Results}
\vskip 0.5cm

The dynamical system given by eqs. (5) and (6) is in general analytically
intractable but can always be solved numerically. Such simulations reveal two
basic dynamic regimes. In the first one, all colonized host are concentrated
in a narrow vicinity of the maximum of the transmission rate
$\tau(x)$.  In this state,
religious variation is 
controlled only by diffusion, i.e. random deviations of the hosts from the 
optimal meme, 
as illustrated in Figure 1a. In
the second regime, frequency-dependent selection on religious memes leads to
the maintenance of religious variation. Maintenance of variation in turn occurs
in two different ways. At equilibrium, the distribution of colonized hosts,
$C(x)$, is either a unimodal function with a 
large 
positive variance (much larger than expected from diffusion alone), as shown in
Figure 1b, or the equilibrium distribution is multimodal, as shown in Figure
1c. Multimodal pattern formation as shown in Figure 1c corresponds to the
emergence of different religions, and hence to religious diversification. 

\begin{figure}[htp]
\epsfig{file=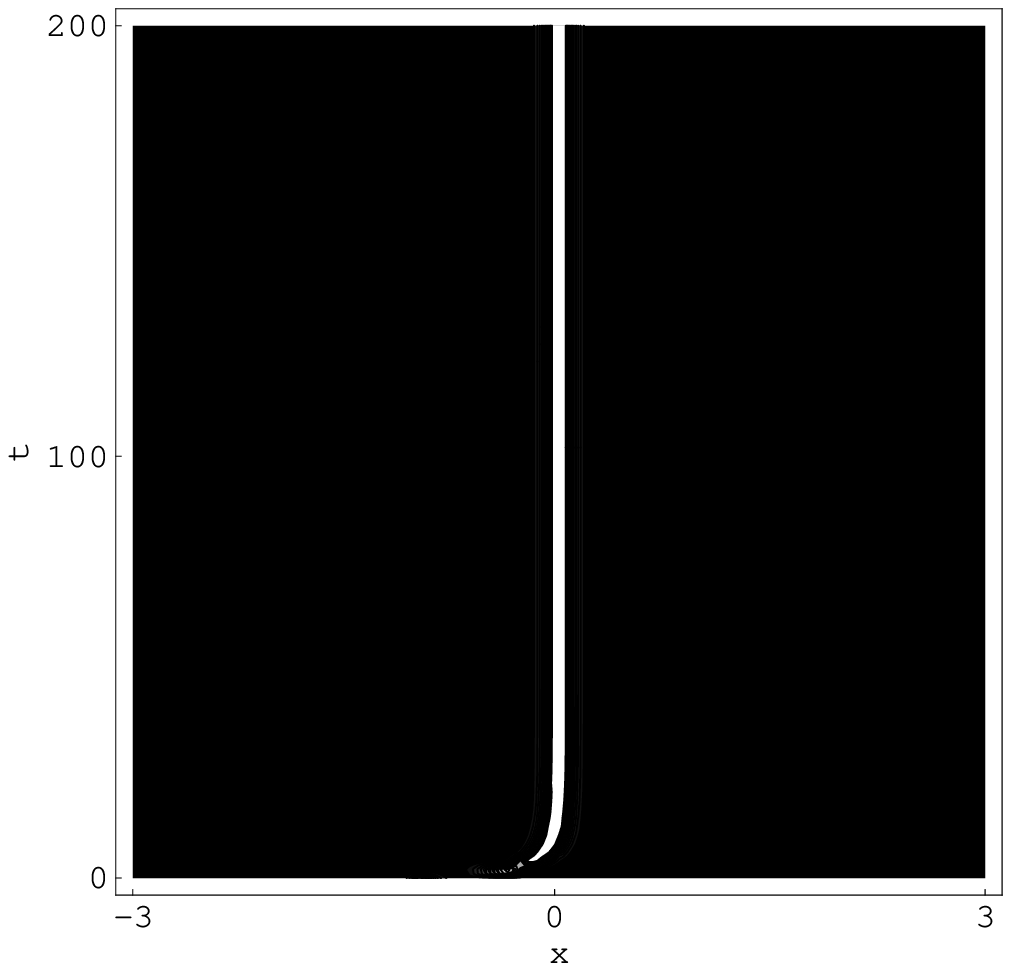,width=0.3\textwidth}
\epsfig{file=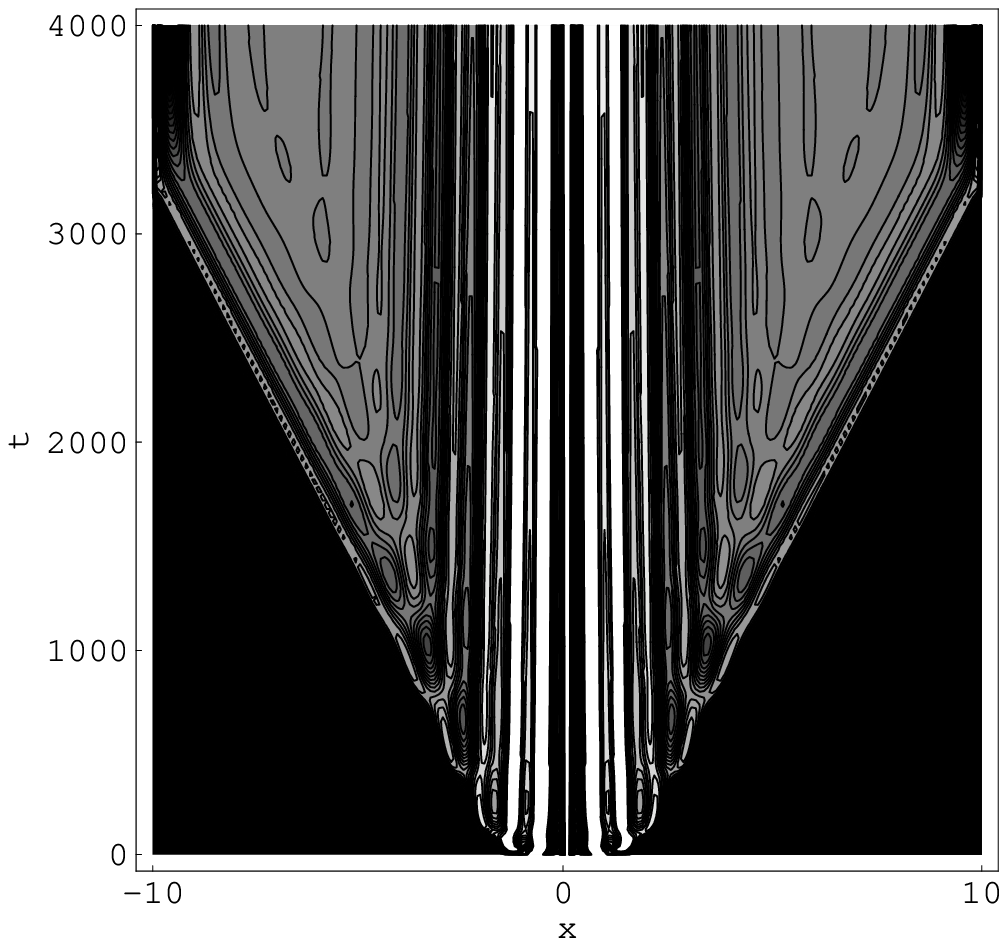,width=0.3\textwidth}
\epsfig{file=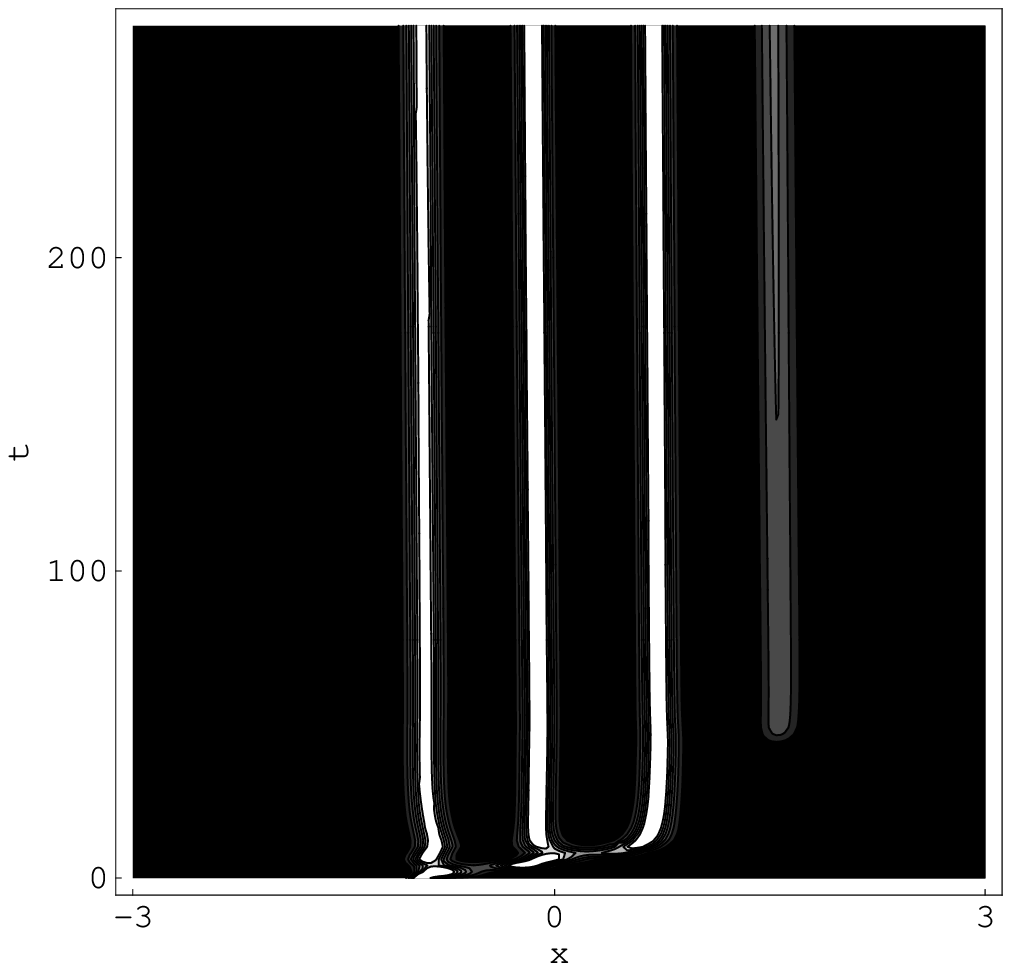,width=0.3\textwidth}
\epsfig{file=./1d.eps,width=0.4\textwidth}

\caption{ \label{Figure 1} Evolution of the religious meme distribution $C(x)$ obtained via numerical solution of
the eqs. (5) and (6). For different values of the parameters
evolution results in (a) no diversification, i.e., a narrow unimodal distribution; b) diversification in the form of a broad unimodal distribution; c) diversification in the form of multimodal distributions, with each mode representing a separate emerging religion. In panel d) the equilibrium distribution $C(x)$ attained in the long-time limit is plotted for the
parameters used in panel a) (solid line), panel b) (dashed line), and panel c) (dot-dashed line). Parameter values were $K_C=K_S=10^4$, $r_C=r_S=1$, $\sigma_m=0.02$, $\tau_0=0.006$ and $\alpha_0=0.003$ for all panels (these parameters were chosen so as to maintain a suitable population size in the individual-based models used for Figure 2). Panel a): $\sigma_\tau = 0.5$, $\sigma_\alpha=1$, $b_\tau=b_\alpha=2$; panel b): $\sigma_\tau = 1$, $\sigma_\alpha=0.5$,
$b_\tau=b_\alpha=2$; panel c): $\sigma_\tau = 1$, $\sigma_\alpha=0.5$, $b_\tau=b_\alpha=3$.}  
\end{figure}

In fact, even if religious diversity ultimately manifests itself in a unimodal
distribution as in Figure 1b, starting from homogenous meme populations
essentially containing only one type of meme, this equilibrium distribution is
reached through a series of ``bifurcations'' into distinct religious
strains, as can be seen in Figure 1b. Over time, different strains give
rise to new strains, a process which eventually fills in the meme space and
results in a unimodal equilibrium distribution. This can be seen more clearly using the individual-based models introduced below (see Figure 2b).

It is worth noting that the case of Gaussian functions $\tau(x)$ and $\alpha(x-y)$ shown in Figure 1b is special in the sense that for these functions, it is possible to find an analytical expression for the equilibrium distribution of colonized memes. Specifically, it is not hard to see that the equilibrium distribution $C(x)$ must satisfy the equation $\int \alpha(x-y)C(y)dy=\tau(x)S$, and if both $\tau(x)$ and $\alpha(x)$ are Gaussian, this equation  has a Gaussian solution $C(x)$ with  width $\sigma=\sqrt{\sigma_\tau^2-\sigma_\alpha^2}$. This is the width of the equilibrium shown in Figure 1b. However, the existence of such an equilibrium is a special property of the ``Gaussian'' case, and finding analytical expressions for equilibrium distributions in cases where the exponents $b_\tau$ and $b_\alpha$ appearing in the functions $\tau(x)$ and $\alpha(x-y)$ are not equal to 2 is in general impossible. In particular, it is in general not true that such equilibrium distributions are unimodal, as the example in Figure 1c shows. 

Whether diversification occurs, and whether diversification manifests itself
in unimodal or multimodal meme distributions, depends on the parameters of the
model. First of all, diversification occurs when $\sigma_\alpha$ is small
enough compared to $\sigma_\tau$. This is revealed by numerical simulations,
and below we will use the framework of adaptive dynamics to provide some
analytical justification for this threshold. Because $\sigma_\alpha$ is a
measure for how fast memes can gain an advantage by being different from
common memes, and $\sigma_\tau$ measures how fast transmissibility decreases
with increasing distance from the optimum $x_0$, this can be roughly
interpreted as diversification in religious memes occurring if the advantage
gained from rarity outweighs the disadvantage due to having lower
transmissibility.

Second, whether diversification, if it occurs, results in unimodal or
multimodal equilibrium distributions depends on the exponents $b_\alpha$ and
$b_\tau$, i.e., on the nature of the functions $\alpha(x-y)$ and
$\tau(x)$. Generally speaking, multimodal meme distributions, and hence
diversification into multiple distinct religious strains, require larger
exponents in these functions. For example, in Figure 1b showing unimodal
diversification, these exponents were set to 2, i.e., both functions
$\alpha(x-y)$ and $\tau(x)$ were of Gaussian form. In this particular case, it
is in fact easy to see that the dynamical system given by eqs. (5) and (6) has an
equilibrium distribution of colonized hosts that is itself Gaussian, and hence
unimodal. However, increasing these exponents to $b_\alpha=3$ and $b_\tau=3$,
as in Figure 1c, results in multimodal equilibrium distributions. Thus,
functions $\alpha(x-y)$ and $\tau(x)$ that fall off less sharply from their
maximum tend to favour multimodal diversification. 

To augment our analysis, we use the mathematical framework of adaptive
dynamics (\cite{metz96, geritz_etal98, dieckmann_law96}), which has proven
itself to
be a very useful tool for identifying various scenarios of evolutionary
diversification and speciation in organismal biology (e.g. \cite{dieck_doe99,
  dieck04}). In this framework, one considers monomorphic resident populations
consisting of a single meme type, and then investigates the fate of rare
mutant memes that appear in the resident population, e.g.  because one of the
hosts colonized by the resident religious meme has slightly changed their
faith and is now host to a slightly altered ``mutant'' meme.  

To do this, we first have to consider the dynamics of monomorphic resident
populations. If the population is monomorphic for memes of trait $x$, the
distribution $C(z)$ is a delta function with total weight $C(x)$ centered at
$x$. Therefore, $A(x)=C(x)$ (eq. (3)). Equations (5) and (6) then become a
system of two ordinary differential equations: 
\begin{align}
\frac{dS}{dt}&=r S\left(1-\frac{S+C(x)}{K}\right)-\tau(x) SC(x)+\alpha_0 C(x)C(x)\\
\frac{d C(x)}{d t}&=r C(x)\left(1-\frac{S+C(x)}{K}\right) + \tau(x) SC(x) -\alpha_0
C(x)C(x). 
\end{align}
It is easy to see that this system has a unique equilibrium 
\begin{align}
(S^*,C^*)=\left(\frac{K\alpha_0}{\alpha_0+\tau(x)},\frac{K\tau(x)}{\alpha_0+\tau(x)}\right)   
\end{align}
at which both $S^*>0$ and $I^*>0$. Moreover, the Jacobian matrix of system
(8), (9) at the equilibrium $(S^*,C^*)$ has two negative eigenvalues, and the
equilibrium is globally stable in the sense that the system will converge to
this equilibrium from any initial condition with both densities $>0$.  

Let's assume that the host population is colonized by a single resident meme
type $x$, and that the resident dynamics given by eqs. (8) and (9) has settled
at its equilibrium $(S^*,C^*)$. This equilibrium constitutes the environment
for a rare mutant meme type $y$ that appears in the host population. If the
mutant is rare, its logistic growth term is determined by the total resident
density $S^*+C^*$, its transmissibility is $\tau(y)$, and its rate of loss of
faith is $A(y)=\alpha_0 \alpha(y-x)C^*$. Therefore, the growth of the
population of hosts colonized by the mutant meme $y$ is  
\begin{align}
\frac{d C(y)}{d t}&=r C(y)(1-\frac{S^*+C^*}{K}) + \tau(y) SC(y)
-\alpha(y-x)C^*C(y) .
\end{align}
The invasion fitness $f(x,y)$ of a rare mutant meme $y$ in the resident $x$
lies at the basis of adaptive dynamics analyses and is defined as the per
capita growth rate of $y$-types, i.e., by the right-hand side of eq. (11)
divided by $C(y)$: 
\begin{align}
f(x,y)=r (1-\frac{S^*+C^*}{K})+\tau(y) S^* -\alpha(y-x)C^*.
\end{align}
According to general theory (\cite{dieckmann_law96}, the adaptive dynamics of
the religious trait $x$ is then given by the selection gradient 
\begin{align}
D(x)=\frac{\partial f(x,y)}{\partial y}\vert_{y=x}=\tau'(x)S^*.
\end{align}
More precisely, the adaptive dynamics of the trait $x$ is 
\begin{align}
\frac{dx}{dt}=\mu D(x),
\end{align}
where $\mu$ is a quantity describing the rate at which resident memes give
rise to mutational variants. 

The analysis of the evolutionary dynamics given by (14) proceeds in two
steps. First one finds stable equilibria of the dynamical system (14), and
then one checks the evolutionary stability of this equilibria, as
follows. Equilibria of (14) are points $x^*$ in meme trait space satisfying
$D(x^*)=0$. In the present case, i.e., with $D(x)$ given by (13), there is
only one such point: the maximum of the function $\tau(x)$, $x^*=x_0$. Dynamic
stability of this so-called singular point is determined by the derivative of
$D(x)$ at the singular point, i.e., by  $dD/dx(x_0)$. In the present case,
this derivative is proportional to the second derivative of $\tau(x)$ at
$x_0$, which is negative. Therefore, the singular point $x^*=x_0$ is a locally
stable attractor for the dynamics (14), and it follows that starting from any
initial resident value $x$, the meme trait will converge to the value $x_0$. 

However, despite this convergence stability the singular point $x_0$ need not
be evolutionarily stable. Evolutionary stability is determined by shape of the
invasion fitness function around the singular point. Note that by definition
of the singular point as a solution of $D(x^*)=0$, the first derivative of the
invasion fitness function is necessarily $0$ at a singular point. Thus,
generically the invasion fitness function either has a maximum or a minimum at
$x_0$. It is shown in adaptive dynamics theory (\cite{geritz_etal98}) that if
$x_0$ is a fitness minimum, this generates the phenomenon of {\it evolutionary
  branching}. Once the resident is at $x_0$, every nearby mutant can
invade. Moreover, two nearby mutants on either side of the singular value
$x_0$ can coexist, leading to meme populations consisting of two coexisting
strains. Finally, in each of these two strains selection favours trait values
lying further away from the singular  point, which means that the two strains
will diverge evolutionarily. The phenomenon of convergence to a singular point
that is a fitness minimum and subsequent emergence and divergence of
coexisting strains is called {\it evolutionary branching}, and the singular
point is called an evolutionary branching point. For example, if we assume
that the two  exponents $b_\alpha$ and $b_\tau$ occurring in the functions
$\alpha(x,y)$ (eq. (4)) and in the function $\tau(x)$ 
are equal to 2,
then one can show that
the singular point $x_0$ is a fitness minimum if  
\begin{align}
\sigma_\alpha<\sigma_\tau.
\end{align}
In particular, evolutionary branching occurs if $\sigma_\alpha$ is small
enough compared to $\sigma_\tau$. 

A single bout of evolutionary branching leads to coexistence of diverging
strains, and it is in principle possible to analyze the (2-dimensional)
adaptive dynamics of these coexisting strains using invasion fitness
functions. In a typical scenario, the two coexisting strains evolve to a new
equilibrium (i.e., a new singular point in 2-dimensional trait space), and
this singular point may or may not be a branching point. If it is, further
bouts of evolutionary diversification occur, resulting in the coexistence of
more than two strains. If it isn't, evolution comes to a halt in a diversified
population consisting of two distinct and coexisting strains. Adaptive
dynamics after diversification into two coexisting strains can in principle be
studied analytically (in a similar way as above, see
e.g. \cite{dieckmann_law96}), but it is also illustrative to study the
evolutionary dynamics in individual-based models. In such models, the various
terms on the right hand side of eqs. (5) and (6) are interpreted as rates at
which birth, death, transmission and loss of faith occur, resulting in a
stochastic model for the evolutionary dynamics. The detailed setup of these
models is described in the Appendix. 
Figure 2 shows different scenarios of evolutionary branching occurring in the
individual-based model.  

\begin{figure}[htp]
\epsfig{file=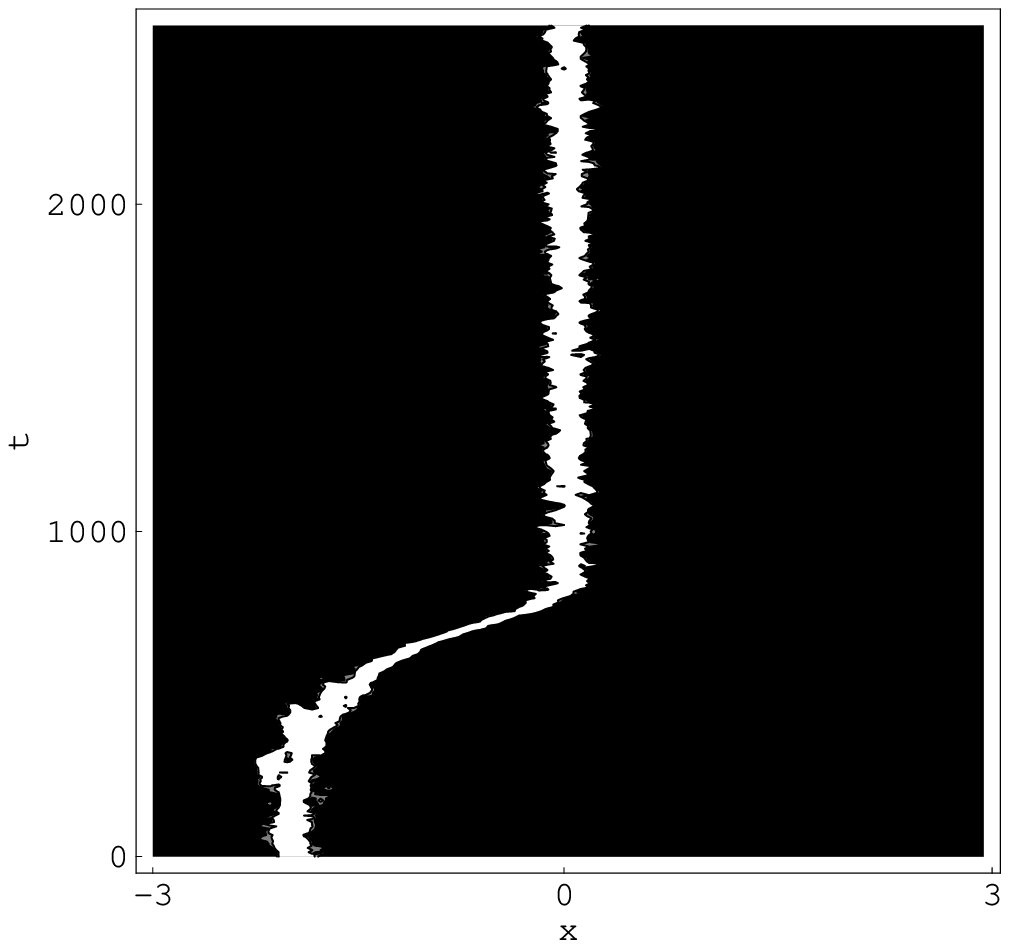,width=0.3\textwidth}
\epsfig{file=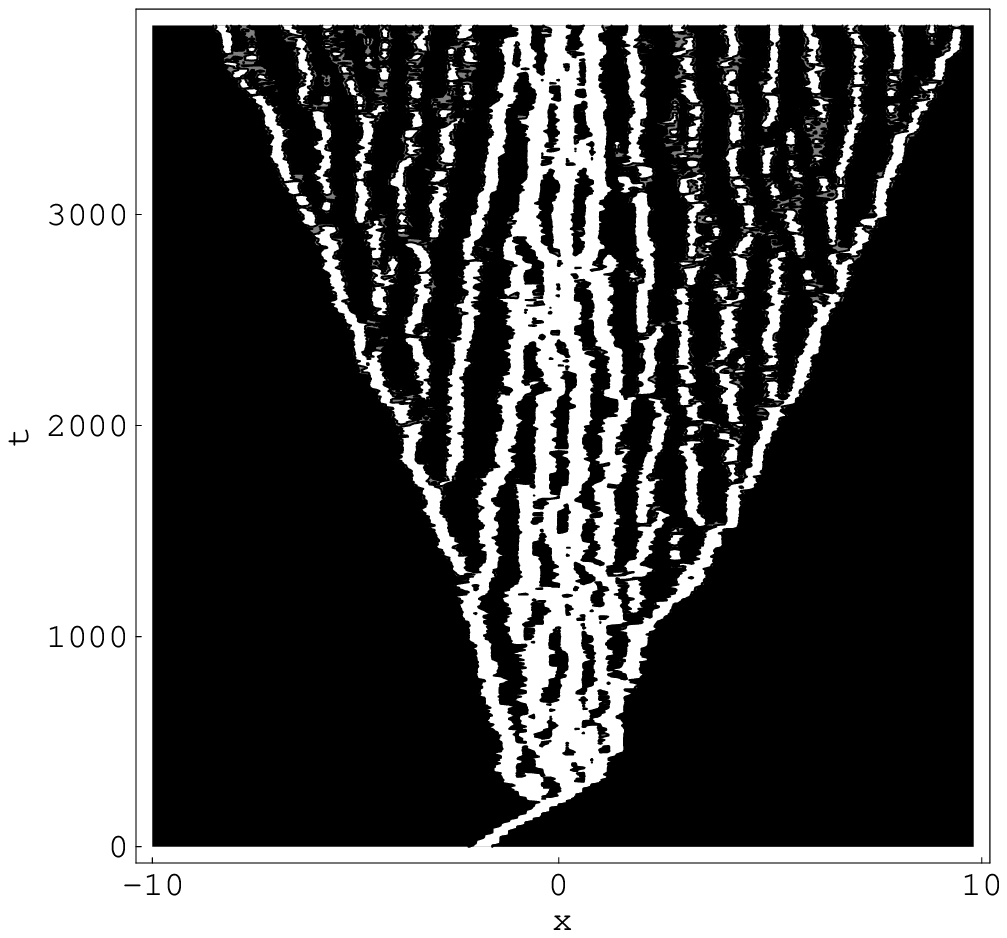,width=0.3\textwidth}
\epsfig{file=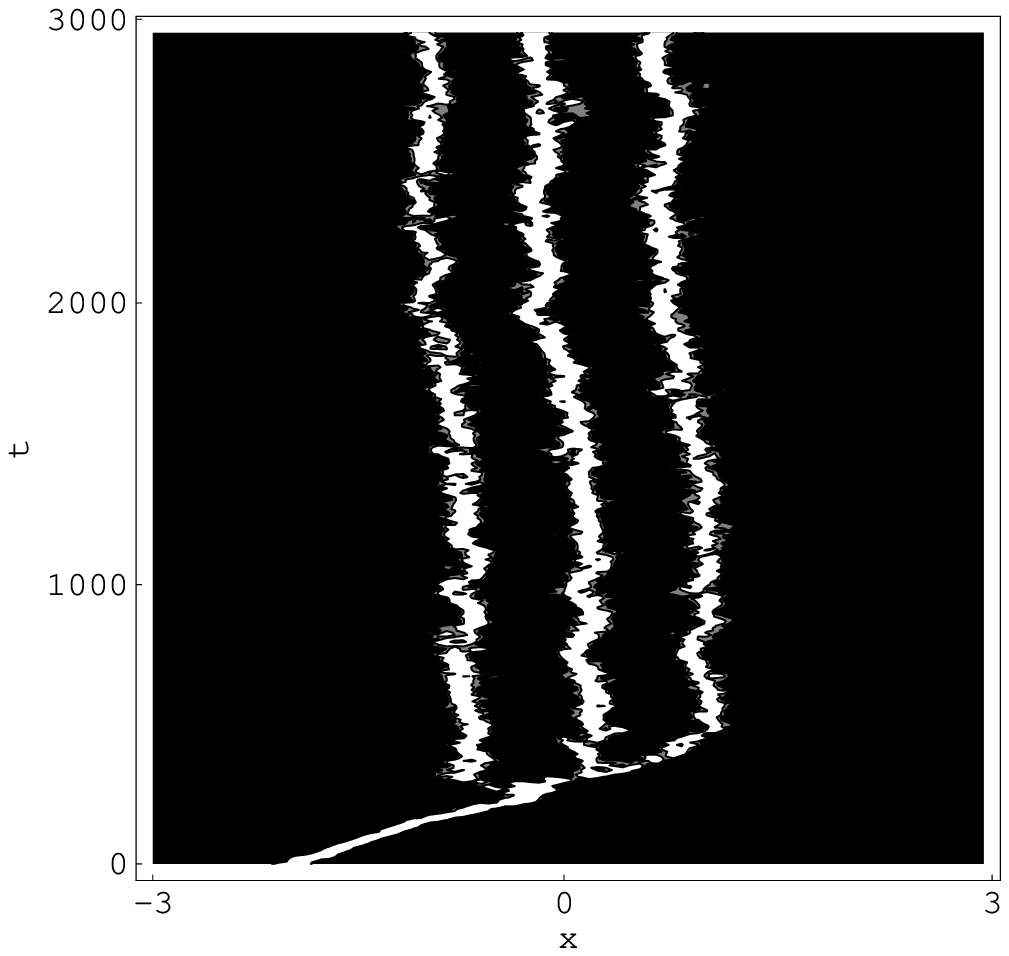,width=0.3\textwidth}
\caption{\label{Figure 2} Evolution of the religious meme distribution $C(x)$ obtained from individual-based models. Parameter values are the same as in the corresponding panels of Figure 1. a) No diversification. b) Continuous sequential branching in the Gaussian case results in an essentially unimodal distribution of religious memes, corresponding to the unimodal equilbrium distribution in Figure 1b. c) Branching stops after two bouts and results in the coexistence of three distinct meme clusters. Note that in contrast to the deterministic model in Fig. 1c, which exhibits five clusters, the individual-based model only shows three clusters due to finite population size.}  

\end{figure}

\vskip 1cm
\noindent{\bf\large Discussion}
\vskip 0.5cm

We have applied the theory of evolutionary diversification to cultural
evolution of religion. Using a simple mathematical model adapted from the epidemiological literature, we have shown that in principle, 
a sufficient ``overcrowding'' of followers of a mainstream religion can lead
to splitting and diversification of religious memes, which manifests itself either as a
broadening of the original religion into a broad ensemble of 
memes, or in splitting into several separate confessions. It is important to realize that this type of diversification occurs not because of spatial separation between different cultures, but because of frequency-dependent selection on religious memes that is mediated by interactions between carriers of different memes. Thus, this type of diversification occurs in situ from a single ancestral religion. The historic
record contains many examples of both types of diversification occurring in our models: emergence
of partially overlapping sects that differ from each other, for example, in the details of the
interpretation of holy texts, and major splits that lead to the emergence of
 separate religion hierarchies, such as between the Catholic,
Protestant,  and Eastern Orthodox churches. The initial branches
often further diversify, for which the fragmentation of the Protestant church, which
peaked in the 19th century, may be a good example. Some of the branches may later merge again, such as in the reunification of the Russian Orthodox Church and the Russian 
Orthodox Church Outside of Russia, which were divided by the 1917 revolution but reunited very recently. 
The sequential branching and later reunification can be observed in the behaviour
of our model (e.g. Fig.1b). A more recent example of the type of cultural evolutionary branching modeled here may be occurring in Papua New Guinea, where \cite{whitehouse95} described the coexistence of a mainstream religious cult with periodically emerging sectarian splinter groups. We think that it might be interesting and fruitful to investigate the driving force for the emergence of such ``modes of religiosity'' (\cite{whitehouse95}) based on the perspectives of frequency-dependent selection and evolutionary branching in religious memes.

Our approach consists of modeling cultural
evolution by considering the cultural memes themselves as the units of selection, rather
than the humans carrying the cultural memes. 
Cultural memes, such as languages
and ideologies, clearly exhibit reproduction and heredity through their
transmission between human hosts. Of course, these memes ultimately need their
human bearers for survival and reproduction 
(for example, a book's content only
comes ``alive'' once the book is read). Just as the survival and reproduction
of symbionts and pathogens is tied to their effects on their hosts, the
evolutionary fate of cultural memes is tied to their impact on human
individuals. And just as viewing individual organisms as hosts of evolving
symbionts or pathogens offers the appropriate perspective for studying the
evolution of those symbionts and pathogens, viewing human individuals as hosts
of evolving memes offers a useful perspective. 

A more traditional approach to cultural evolution consists of viewing different human populations as carrying different cultural memes, and of investigating competition between such human populations. In the language of host-pathogen models, this would correspond to considering different host populations carrying different pathogens and asking which of the host populations can outcompete the other. Because in this perspective success is based on characteristics imparted or imposed by the pathogen on a group of hosts, this perspective is akin to group selection. In contrast, studying the evolution of pathogens or the evolution of cultural memes in a single host population is based on individual selection in the pathogens or cultural memes. In our model, this difference to traditional approaches is reflected in the fact that the religious trait $x$ does not affect survival and reproduction in the host. Thus, selection on memes is not mediated by differential viability or reproductive success in the host population, but by the fact that different memes have different rates of being transmitted to susceptible hosts, and different rates of being lost from colonized hosts. Loss of memes is frequency-dependent, because the rate of loss depends on overcrowding, which is the driving force of diversification. 

Just as in host-pathogen or host-symbiont systems, coevolution between humans and culture
may be very important, and one can easily envisage many extensions of the model presented here to more complicated scenarios, in which the effects of culture on individual hosts as well as on the demographics of entire host populations are described in more mechanistic detail (e.g. in terms of propensity of host reproduction and sacrifice), and in which genetic evolution in the host occurs as a response to constraints imposed by cultural content, which in turn changes cultural opportunities. We believe that the perspective of cultural memes
as the evolutionary unit will be very useful for such studies. We also think that this perspective serves to objectify and
quantify the significance of cultural content, such as religion. Cultural content is best
viewed not as fixed and pre-existing, but as evolving due to its effect on
human individuals, who ultimately decide whether to accept or reject such
content and how vigorously to spread it upon acceptance.


\newpage
\bibliographystyle{rel}
\bibliography{relig}
\newpage
\vskip 1cm
\noindent{\bf\large Appendix}
\vskip 0.5cm

To construct individual-based stochastic models that correspond to the deterministic model given by eqs. (5) and (6), we have to distinguish the different types of events that can occur at the level of individuals: birth, death, loss of faith, and transmission of faith. Each of these events occur at certain rates. For example, all host individuals have a per capita birth rate $r_S=r_C$, so that the total birth rate of susceptible hosts, $B_S$, is $r_SS$, and the total birth rate of colonized hosts, $B_C$, is $r_CC$, where $S$ and $C$ are the number of susceptible and colonized hosts present in the population at any given time (note that in contrast to eqs. (5) and (6), where $S$ and $C$ are population densities, and hence real numbers, in the indiviudal-based models $S$ and $C$ are integers). For both susceptible and colonized host individuals, the per capita death rate is $r_SSC/K_S=r_ISC/K_I$, and total death rates $D_S$ and $D_C$ for susceptible and colnized individuals are $r_SS^2C/K_S$ and $r_ISC^2/K_I$, respectively. For a host colonized by religious meme $x$, the per capita rate at which this meme is transmitted to susceptible hosts is $\tau(x)S$, where $\tau(x)$ is the transmission function (7). The total rate of transmission, $T$, is therefore $\sum_i\tau(x)S$, where the sum runs over all colonized hosts. Finally, the per capita rate of loss of religion of host individuals colonized by religious meme $x$ is given by $c(A(x))=A(X)$ (eq. (3)), so that the total rate of loss, $L$, is  $\sum_iA(x)$.

The individual-based model is implemented as follows. At any given time $t$, all individual rates as well as the total rates $B_S$, $B_C$, $D_S$, $D_C$, $T$ and $L$ are calculated as described above. Then the type of event that occurs next, birth or death of a susceptible or a colonized host, transmission of a religious meme, or loss of a meme, is chosen with probabilities proportional to the total rates for these events (i.e., with probabilities $B_S/E$, etc., where $E=B_S+B_C+D_S+D_C+T+L$is the total event rate). For the event chosen, the individual to perform this event is chosen with probabilities proportional to the individual rates for the chosen event. For example, if loss of faith is the chosen event, individual $i$ is chosen to lose faith with probability $A(x)/L$, where $x$ is the religious meme of individual $i$. This individual is then removed from the population of colonized hosts, and the number of susceptible hosts is augmented by 1. Similarly, if the chosen event is transmission, individual $i$ among the colonized hosts is chosen for transmission with probability $\tau(x)S/T$, where $x$ is again the meme of individual $i$. Individuals for birth and death events are chosen analogously. If an individual dies it is removed from the population (and the numbers $S$ or $C$ are updated accordingly). If a susceptible individual gives birth the number $S$ is augmented by 1. If a colonized individual with meme $x$ gives birth, a new colonized host is added to the population carrying a meme $x'$ that is chosen from a normal distribution with mean the parental meme $x$ and a certain (small) width $\sigma_{m}$. This reflects ``mutation'' during transmission of religious memes from parent to offspring.

Performing one individual event in the manner described above completes one computational step in the individual-based model, which advances the system from time $t$ to time $t+\Delta t$ in real time. To make the translation from discrete computational steps to continuous real time, $\Delta t$ is drawn from an exponential probability distribution with mean $1/E$, where $E$ is the total event rate. Thus, if the total event rate $E$ is high, the time lapse $\Delta t$ between one event and the next is small, and vice versa if the total event rate is low. Starting from some initial population containing $S_0$ susceptible hosts and $C_0$ colonized hosts with memes $x_1^0,...,x_{N_0}^0$ at time $0$, iteration of the computational steps described above generates the stochastic evolutionary dynamics of a finite population in continuous time.

\end{document}